\documentclass[11pt,a4paper]{article}
\pdfoutput=1
\usepackage{jcappub}
\usepackage{color}
\input{colordvi.tex}
\usepackage{graphicx}
\usepackage{float}
\usepackage{wrapfig}
\usepackage{bm}
\usepackage{amsmath,amssymb,ascmac}
\usepackage{subfigure}
\usepackage{verbatim}
\usepackage{makeidx}

\title{Secondary isocurvature perturbations from acoustic reheating}
\author{Atsuhisa Ota}
\author{and Masahide Yamaguchi}

\affiliation{Department of Physics, Tokyo Institute of Technology,\\
Tokyo 152-8551, Japan}
\emailAdd{a.ota@th.phys.titech.ac.jp}
\emailAdd{gucci@phys.titech.ac.jp}

\abstract{The superhorizon (iso)curvature perturbations are
conserved if the following conditions are satisfied: (i) (each) non
adiabatic pressure perturbation is zero, (ii) the gradient terms are
ignored, that is, at the leading order of the gradient expansion (iii)
(each) total energy momentum tensor is conserved. We consider the case
with the violation of the last two requirements and discuss the
generation of secondary isocurvature perturbations during the late time
universe. Second order gradient terms are not necessarily ignored even
if we are interested in the long wavelength modes because of the
convolutions which may pick products of short wavelength perturbations
up. We then introduce second order conserved quantities on superhorizon
scales under the conditions (i) and (iii) even in the presence of the
gradient terms by employing the full second order cosmological
perturbation theory.  We also discuss the violation of the condition
(iii), that is, the energy momentum tensor is conserved for the total
system but not for each component fluid.  As an example, we explicitly
evaluate second order heat conduction between baryons and photons due to
the weak Compton scattering, which dominates the period just before
recombination.  We show that such secondary effects can be recast into
the isocurvature perturbations on superhorizon scales if the local type
primordial non Gaussianity exists a priori.}


\begin{document}
\maketitle

\section{Introduction}

The conserved quantities on superhorizon scales play an important
role in inflationary Universe because they connect the primordial
perturbations generated during inflation with those at the late
Universe.  Even though most of characteristic signals of the early
universe are washed away due to the thermalization processes, they keep
the statistics of the primordial density fluctuations, which enables us
to reveal the details of inflationary models.  We usually evaluate such
quantities when they exit horizons during inflation and consider them as
initial conditions of the hot Big Bang universe.  The curvature
perturbation on the uniform density slice $\zeta$ is one of typical
examples of such conserved
quantities~\cite{Malik:2003mv,Lyth:2003im,Lyth:2004gb}.  Suppose the
total energy momentum tensor is conserved and we drop the gradient
terms, it is well-known that $\zeta$ is conserved even at nonlinear
order when there are no non adiabatic pressure perturbations.  We can
also define the curvature perturbations $\zeta_{\alpha}$ on
$\alpha$-fluid uniform density slice, where $\alpha=\nu,~b,~c$
represents neutrino, baryon, or cold dark matter (CDM) while $\gamma$
the photon fluid.  Then, the isocurvature perturbations are introduced
as $S_{\alpha \gamma}\equiv 3(\zeta_{\alpha}-\zeta_{\gamma})$.  It
should be noticed that $S_{\alpha \gamma}$ are also conserved at the
leading order of the gradient expansion if the energy momentum tensors
of $\alpha$- and $\gamma$-fluids are conserved, respectively.  The
conservation law of the total energy momentum tensor is universal so
that the conservation laws of the curvature perturbations have been also
considered to be robust as long as the other conditions are satisfied.

In this paper, we revisit the above two assumptions for the conservation
laws of $\zeta_{\alpha}$: ignoring the gradient terms and the
conservation laws of the energy momentum tensors.  First, we point out
that, at nonlinear order, we cannot justify to drop the gradient terms
even when we consider the long wavelength modes; convolutions in Fourier
space can pick up products of short wavelength modes, which might be
significant.  As a result, the total curvature perturbations might not
be conserved at nonlinear order even without non-adiabatic pressure
perturbations.  We then newly introduce a second order conserved
quantity in the presence of gradient terms.  Second, we discuss energy
transfer among components, that still conserves the total energy
momentum but violates each one. This would lead to the evolution of
superhorizon isocurvature perturbations. The typical example of the
above process is acoustic reheating of the photon-baryon
plasma~\cite{Jeong:2014gna,Nakama:2014vla,Naruko:2015pva}.  The short
wavelength temperature fluctuations of the cosmic microwave background
(CMB) are significantly damping due to imperfectness of the
photon-baryon fluid, which produces the second order entropy production
and the second order energy transfer between the photons and the
baryons.  These processes actually happen inside each diffusion scale;
the secondary effects fluctuate on scales larger than those of corse
graining.  The distant patches are not necessarily reheated
homogeneously if there exist three or four-point correlations of
primordial density perturbations a priori~\cite{Naruko:2015pva}.  They
are comparable to the non gradient terms at second order because the
convolutions pick heat conduction and shear viscosity on small scales up.
We investigate these diffusion effects in detail by employing the
nonlinear cosmological perturbation theory, which enables us to follow
the evolution of the photon distribution function directly.

We organize this paper as follows.  First of all, we explain our set up
for the second order perturbation theory in section~\ref{nonlinearpert}.
Then, we discuss the non conservation of the curvature perturbations in
the presence of gradient terms and introduce a new conserved quantity in
section~\ref{EMTevol}.  Section~\ref{evophoton} is devoted to describe
the actual time evolution of the photon baryon plasma due to the weak
Compton scattering.  We comment on several definitions for the
isocurvature perturbations during non-equilibrium periods in
section~\ref{sec:entropy}.  In the final section, we summarize our
conclusions and describe future prospects related to the present
results.

\section{Set up for second order perturbation theory}\label{nonlinearpert}

We need to perturb both the gravity and the matter sectors up to
nonlinear order.  Here, let us first define the nonlinear metric
perturbations.

\subsection{The metric perturbations}\label{section:metric}

We start with writing the spacetime metric in the following 3+1 form: 
\begin{align}
ds^2&=-\mathcal N^2 d\eta^2 + \gamma_{ij}(\beta^i d\eta + dx^i)(\beta^j d\eta + dx^j)\notag \\
&=(-\mathcal N^2+\beta_k\beta^k) d\eta^2 +2\beta_i dx^i d\eta+ \gamma_{ij} dx^i dx^j.\label{def:metric}
\end{align}
In other words, each component can be written as
\begin{align}
g_{\mu\nu}&=\left(\begin{array}{cc}-\mathcal N^2+\beta_k\beta^k & \beta_j \\\beta_i & \gamma_{ij}\end{array}\right),\label{def:comp}
\end{align}
where $\mathcal N$ and $\beta_i$ are the lapse and the shift, respectively.
$\gamma_{ij}$ is the spatial metric. 
Let us consider nonlinear scalar perturbations introduced as
\begin{align}
\mathcal N^2&=a^2 e^{2{A}},\\
\beta_i&=a^2e^{{D}}\partial_i e^{B},\label{def:shift}\\
\gamma_{ij}&=a^2e^{2{D}}\delta_{ij}\label{def:gamma},
\end{align}
where $a$ is the scale factor, and we have fixed only the spacial coordinate by vanishing the anisotropic part of $\gamma_{ij}$.
The nonlinear metric perturbations can be expanded as $X \equiv \sum_{n=1}
X^{(n)}$ for $X=A,B$ and $D$ with $n$ being the order in primordial perturbations.  
Note that the conformal Newtonian, the uniform density, the spatially
flat and the velocity orthogonal isotropic gauges~(comoving gauge)
are mutually transformed by changing only the time slice.
Here, we ignore the vector and the tensor perturbations for simplicity.
This would be justified if the primordial vector perturbations and the primordial tensor ones are subdominant compared to the second order scalar ones.
We include the curvature perturbation $D$ in Eq.~(\ref{def:shift}) to simplify the inverse matrix in the following discussions.
The inverse matrixes for the induced metric and the shift vector are written as 
\begin{align}
\gamma^{ij}&=a^{-2}e^{-2{D}}\delta_{ij},\\
\beta^i&=e^{-{D}}\partial_ie^{B}.
\end{align}
Then, we obtain
\begin{align}
\beta^k\beta_k&=a^2\partial e^{{B}}\partial e^{{B}},\\
-\mathcal N^2+\beta_k\beta^k&=-a^2e^{2{A}}+a^2\partial e^{{B}}\partial e^{{B}},\label{def:lapse}
\end{align}
where we write as $\partial X\partial Y\equiv \partial_i X\partial_i Y$ for notational simplicity.
Eqs.~(\ref{def:shift}), (\ref{def:gamma}) and (\ref{def:lapse}) yield
\begin{align}
g_{00}&=-a^2e^{2{A}}+a^2e^{2{B}}(\partial {{B}})^2,\\
g_{0i}&=a^2e^{{D}+{B}}\partial_i {B},\\
g_{ij}&=a^2e^{2{D}}\delta_{ij}.
\end{align}
The inverse matrix of Eq.~(\ref{def:comp}) is well known:
\begin{align}
g^{\mu\nu}&=\left(\begin{array}{cc}-\frac{1}{\mathcal N^2} & \frac{\beta^j}{\mathcal N^2} \\\frac{\beta^i}{\mathcal N^2} & \gamma^{ij}-\frac{\beta^i\beta^j}{\mathcal N^2}\end{array}\right).
\end{align}
Then, each component of the inverse matrix can be obtained as

\begin{align}
g^{00}&=-a^{-2}e^{-2{A}},\\
g^{0i}&=a^{-2}e^{-2{A}-{D}+{B}}\partial_i{B},\\
g^{ij}&=a^{-2}e^{-2{D}}\delta^{ij}-a^{-2}e^{-2{A}-2{D}+2{B}}\partial_i{{B}}\partial_j{{B}}.
\end{align}
The determinant of $g_{\mu\nu}$ can be also evaluated as
\begin{align}
\sqrt{-g}=\mathcal N\sqrt{\gamma}=a^4 e^{{A}+3{D}}.\label{def:det}
\end{align}

\subsection{The Christoffel symbols at second order}

Here and hereafter we consider only the perturbations up to second order.
Up to second order, each component of the metric tensor can be rewritten as

\begin{align}
g_{00}&=-a^2e^{2{A}}+a^2(\partial {{B}})^2,\\
g_{0i}&=a^2e^{{D}+{B}}\partial_i {B},\\
g_{ij}&=a^2e^{2{D}}\delta_{ij},
\end{align}
and the inverse matrix components are
\begin{align}
g^{00}&=-a^{-2}e^{-2{A}},\\
g^{0i}&=a^{-2}e^{-2{A}-{D}+{B}}\partial_i{B},\\
g^{ij}&=a^{-2}e^{-2{D}}\delta^{ij}-a^{-2}\partial_i{{B}}\partial_j{{B}}.
\end{align}
Let us evaluate the Christoffel symbol
\begin{align}
\Gamma^\mu{}_{\nu\rho}\equiv \frac12g^{\mu\alpha}\left(\partial_\rho g_{\alpha\nu}+\partial_\nu g_{\alpha\rho}-\partial_\alpha g_{\nu\rho} \right).
\end{align}
Each component of the symbols can be calculated as
\begin{align}
\Gamma^0{}_{00}=&\mathcal H+{A}'+\mathcal H (\partial {B})^2+ \partial A\partial B,\\
\Gamma^0{}_{0i}=&\partial_i{A}+e^{-2{A}+{D}+B}(\mathcal H +{D}' )\partial_i {B}-\frac12\partial_i(\partial{B})^2,\\
\Gamma^0{}_{ij}=&
\frac12 \left[\partial_i {B}\partial_j{D}
+ \partial_j {B}\partial_i{D}\right]
-e^{-2{A}+{D}+{B}}\partial_i\partial_j{B}
\notag \\
&-\partial_i{B}\partial_j{B}
+e^{-2{A}+2{D}}\delta_{ij}\left[
\mathcal H+{D}'-\partial B\partial D\right],\\
\Gamma^{i}{}_{00}=&
e^{-{D}+{B}}(\mathcal H\partial_iB + \partial_iB')
+(-A'+D'+B')\partial_i B\notag\\
&+e^{-2{D}+2{A}}\partial_i{A}-\frac12\partial_i (\partial{B})^2,\\
\Gamma^i{}_{0j}=&(\mathcal H+{D}')\delta_{ij}-\partial_i {B}\partial_j {A}-\mathcal H\partial_i {B}\partial_j {B}\notag \\
& -\frac12(\partial_i{D}\partial_j {B}-\partial_j{D}\partial_i {B})  ,\\
\Gamma^i{}_{jk}=&-\partial_i{D}\delta_{jk}+\partial_k{D}\delta_{ij}+\partial_j{D}\delta_{ik}+(\partial_i{B})\partial_j\partial_k{B}\notag \\
&
-e^{-2{A}+{D}+{B}}(\mathcal H+{D}')\delta_{jk}\partial_i{B}.
\end{align}

\section{Conserved quantity at second order}
\label{EMTevol}

In this section we show the conservation laws of the curvature
perturbations and discuss the gradient corrections by full
consideration of second order perturbation theory.

\subsection{Divergence of the energy momentum tensor}

Let $T^{(\alpha)\mu\nu}$ be energy momentum tensors of $\alpha$-fluid.
Assuming the conservation of the energy momentum tensor for each fluid
component
\begin{align}
\nabla_\mu T^{(\alpha)\mu\nu}=0,\label{cons:colless}
\end{align}
the curvature perturbations on $\alpha$-fluid uniform density slice
\begin{align}
\zeta_\alpha\equiv D+\frac13\int^{\rho(\eta,\mathbf x)}_{\rho_{\rm rf}(\eta)} \frac{d \rho_\alpha}{\rho_\alpha+P_\alpha},\label{defzetatotal}
\end{align}
are conserved as long as non-adiabatic pressure perturbations and the
gradient terms are negligible~\cite{Lyth:2004gb}. 
Let us first take a closer look at the above theorem. In this
section, we do not specify a fluid component explicitly and drop
the symbols from expressions.  The time component of the covariant
divergence can be given as
\begin{align}
\nabla_\mu T^\mu{}_0=&\partial_\mu T^\mu{}_0 +T^\mu{}_0 \partial_\mu \ln\sqrt{-g}-\Gamma^\alpha{}_{\mu 0}T^\mu{}_\alpha.\label{tenkaicovemt}
\end{align}
Note that only a spatial gradient term in
\begin{align}
\partial_\mu T^\mu{}_0=\partial_0 T^0{}_0 + \partial_i T^i{}_0,\label{partderivs}
\end{align}
is negligible on superhorizon scales.  The other gradient terms arising
in products of the linear perturbations cannot be dropped without their
concrete evaluations since they may have significant contributions on
small scales through convolutions in Fourier space.  On the other
hand, from Eq.~(\ref{def:det}), the second term in
Eq.~(\ref{tenkaicovemt}) can be easily evaluated as
\begin{align}
T^\mu{}_0 \partial_\mu \ln\sqrt{-g}=(4\mathcal H+{A}'+3{D}')T^0{}_0+T^i{}_0\partial_i({A}+3{D}).\label{volumefac}
\end{align}
The term with the Christoffel symbol in Eq.~(\ref{tenkaicovemt}) is decomposed into 4 parts:
\begin{align}
\Gamma^\alpha{}_{\mu 0}T^\mu{}_\alpha&=\Gamma^0{}_{0 0}T^0{}_0+\Gamma^0{}_{i 0}T^i{}_0+\Gamma^i{}_{0 0}T^0{}_i+\Gamma^i{}_{j 0}T^j{}_i.
\end{align}
Each part can be easily calculated as
\begin{align}
\Gamma^0{}_{0 0}T^0{}_0=&(\mathcal H+{A}')T^0{}_0+\mathcal H(\partial{B})^2T^0{}_0+(\partial A\partial B) T^0{}_0, \\
\Gamma^0{}_{i 0}T^i{}_0=&\mathcal H \partial_i{B} T^i{}_{0}+\partial_i{A} T^i{}_{0}, \\
\Gamma^i{}_{0 0}T^0{}_i=&
\mathcal H\partial_i BT^0{}_i+ \partial_i B'T^0{}_i + \partial_iAT^0{}_i\\
\Gamma^i{}_{j 0}T^j{}_i=&
3P(\mathcal H +{D}') -P(\partial{B}\partial{A})-P\mathcal H(\partial {B})^2,\label{Gamma0ijsubs}
\end{align}
where we have decomposed $T^i{}_j$ into the trace part (that is, the
pressure part) and the traceless part (the anisotropic pressure part),
\begin{align}
 T^i{}_j = P \delta^i{}_j + \widetilde{T}^i{}_j
\end{align}
with $\widetilde{T}^i{}_i = 0$.
Note that the anisotropic pressure is at least first order quantity, which would be included in the cubic order terms above; therefore, only the isotropic pressure
arises in Eq.~(\ref{Gamma0ijsubs}).  At linear order, the following
relation is useful:
\begin{align}
T^i{}_0+T^0{}_i=-\partial_iB(T^0{}_0 - P).\label{christoffelpart}
\end{align}
Then, using Eqs.~(\ref{partderivs}), (\ref{volumefac}) and (\ref{christoffelpart}), we finally obtain

\begin{align}
\nabla_\mu T^\mu{}_0=&\partial_\mu T^\mu{}_0+3(\mathcal H+{D}')(T^0{}_0-P)-(T^0{}_0-P)\partial{B}\partial({A}+3D)-T^0{}_i\partial_i(A+3D+B').
\label{emtdiv0com}
\end{align}
In most of the previous literatures where perfect fluid
approximations are assumed, the gradient terms are automatically
dropped. On the other hand, in our case, only the second term in
Eq.~(\ref{partderivs}) is negligible, and products of the linear
perturbations cannot be necessarily dropped.  Let us introduce the
energy density $\rho$ and the momentum transfer $q$ as
\begin{align}
\rho&\equiv -T^0{}_0,\\
\partial_i q&\equiv \frac{ T^0{}_i}{\rho+P}.
\end{align}
Then, (\ref{emtdiv0com}) can be recast into
\begin{align}
-\frac{1}{3(\rho+P)}\nabla_\mu T^\mu{}_0&=\mathcal H + D'+\frac{\rho'}{3(\rho+P)}-\frac13\partial{B}\partial({A}+3D)
+\frac13\partial q\partial(A+3D+B').
\label{emtdiv0com3}
\end{align}

Note that we have not taken the specific time slice other than the spacial coordinate; therefore Eq.~(\ref{emtdiv0com}) is useful for conformal Newtonian~($B=0$), uniform density~($\delta \rho=0$), spatially flat~($D=0$) or velocity orthogonal isotropic gauges~($q=0$), respectively.\\

\subsection{Gradient corrections}

We are now ready to discuss the superhorizon conserved quantities in the presence of gradient terms. 
From Eqs.~(\ref{cons:colless}), (\ref{defzetatotal}), and~(\ref{emtdiv0com3}), we immediately obtain

\begin{align}
\zeta_\alpha'=&\frac13\partial{B}\partial({A}+3D)
-\frac13\partial q_\alpha \partial(A+3D+B').
\label{non:cons:zeta}
\end{align}
Eq.~(\ref{non:cons:zeta}) apparently shows that $\zeta_{\alpha}$ is not
conserved in the presence of second order gradient terms.  Note that we
cannot simply ignore the RHS even for long wavelength modes as we
already mentioned.\\

As explained in section~\ref{section:metric}, the spacial coordinate is
already fixed; the residual linear gauge freedom is given by a shift of
the time coordinate
\begin{align}
\eta &\to \eta+\alpha.\label{alphadefgauge}
\end{align}
Here, it should be noticed that the source term is composed of the
products of linear perturbations; therefore, we only consider the linear
gauge transformation here.  In response to the above transformation, the
metric perturbations obey the following transformation
laws~\cite{Ma:1995ey}:
\begin{align}
A&= \tilde A-\alpha'-\mathcal H\alpha,\\
B&= \tilde B+ \alpha,\\
D&= \tilde D-\mathcal H\alpha.
\end{align}
On the other hand, the energy density, the pressure and the momentum transfer transform as
\begin{align}
\delta \rho &= \delta\tilde \rho -\alpha\rho^{(0)}{}',\\
\delta P &= \delta\tilde P -\alpha P^{(0)}{}',\\
q &= \tilde q + \alpha.\label{gt:q}
\end{align}
Then, we find
\begin{align}
A+3D+B'&=\tilde A+3\tilde D+\tilde B'-4\mathcal H \alpha.\label{gauge:a3dbp}
\end{align}
Eqs.~(\ref{non:cons:zeta}) and (\ref{gauge:a3dbp}) motivate us to move on to the gauge which satisfies the following relation: 
\begin{align}
A+3D+B'=0.\label{mygauge}
\end{align}
This condition is useful since the fluid components and metric perturbations decouple in the covariant derivative of the energy momentum tensor, and gauge fixing is complete from Eq.~(\ref{gauge:a3dbp}).
In this gauge, we find following quantities are conserved:
\begin{align}
\xi_{\alpha} \equiv D + \frac16\partial{B}\partial B+\frac13\int^{\rho(\eta,\mathbf x)}_{\rho_{\rm rf}(\eta)} \frac{d \rho_\alpha}{\rho_\alpha+P_\alpha}.\label{cons:grad:second}
\end{align}
Note that $\xi_{\alpha}\to \zeta_{\alpha}$ if we ignore the gradient
term. 
We define the isocurvature perturbations in terms of
$\xi_{\alpha}$ in the similar way:
\begin{align}
S_{\alpha \gamma}=3(\xi_{\alpha}-\xi_{\gamma}),\label{def:iso:xi}
\end{align}
which are also conserved if the energy momentum tensors are conserved
and non-adiabatic pressure perturbations are absent. Thus the
curvature perturbations on the uniform density slice are no more
conserved in the presence of gradient terms.  Instead, we introduced
another conserved quantity $\xi$ at second order.  $\xi$ is no more the
curvature perturbation on the uniform density slice since we moved to
another specific time slicing.  In the next section, we consider
the time evolution of $\xi$ in the presence of a collision
process.

\section{Energy transfer and time evolution of the isocurvature perturbations}\label{evophoton}

\subsection{The local Minkowski frame for collision processes}

Here, we discuss the collision processes for the weak Compton
scattering, which are described by the quantum electrodynamics~(QED) in
the local Minkowski coordinate.  To relate the local frame with the
global one defined in Eq.~(\ref{def:metric}), let us consider the
following coordinate
transformations~\cite{Pitrou:2007jy,Naruko:2013aaa}:

\begin{align}
g_{\mu\nu}=\eta_{\bar\alpha\bar\beta}e^{\bar\alpha}{}_\mu e^{\bar\beta}{}_\nu,
\end{align}
where each vierbein is defined as
\begin{align}
e^{\bar 0}{}_0&=ae^A,\\
e^{\bar 0}{}_i&=0,\\
e^{\bar a}{}_0&=ae^B\partial_{\bar a}B,\\
e^{\bar a}{}_i&=ae^{D}\delta_{\bar ai}.
\end{align}
For the inverse matrix, the coordinate transformation becomes
\begin{align}
g^{\mu\nu}=e^\mu{}_{\bar\alpha} e^\nu{}_{\bar\beta}\eta^{\bar\alpha\bar\beta},
\end{align}
where we have introduced
\begin{align}
e^{0}{}_{\bar 0}&=a^{-1}e^{-A},\\
e^{0}{}_{\bar a}&=0,\\
e^{i}{}_{\bar 0}&=-a^{-1}e^{-A-D+B}\partial_{i}B,\\
e^{i}{}_{\bar a}&=a^{-1}e^{-D}\delta_{i\bar a}.
\end{align}

Next, let us consider the physical momentum $\tilde p_{\bar\alpha}$ of a particle in the local Minkowski frame.
The momentum satisfies 
\begin{align}
\tilde p_{\bar\alpha}\tilde p^{\bar\alpha}=\eta^{\bar\alpha\bar\beta}\tilde p_{\bar\alpha}\tilde p_{\bar\beta}=\eta_{\bar\alpha\bar\beta}\tilde p^{\bar\alpha}\tilde p^{\bar\beta}=-m^2,
\end{align}
where $m$ is the mass of the particle.
The evolution of the photon momentum in the expanding universe is written as
\begin{align}
\tilde p^{\bar\alpha}\propto \frac{1}{a}.
\end{align}
Then, it would be more convenient to introduce the comoving momentum so as to subtract the background spacetime evolution.
For this purpose, we define the comoving momentum of the conformal flat coordinate as
\begin{align}
p^{\bar\alpha}\equiv a \tilde p^{\bar\alpha}.
\end{align}
The energy and the spacial direction of the photon are also introduced as
\begin{align}
p&\equiv p^{\bar 0},\label{pdef}\\
n^{\bar a}&\equiv \frac{p^{\bar a}}{p}.
\end{align}
Then we can write the conjugate momentum,
$P^\mu=e^\mu{}_{\bar\alpha}\tilde p^{\bar\alpha}$, associated with
the spacial coordinate by using $p$ and $n^i$ as
\begin{align}
P^0&=\frac{\tilde p^{\bar 0}}{ae^A}=\frac{p}{a^2e^A},\label{P0def}\\
P^i&=\frac{p}{a^2e^D}(n^i-e^{B-A}\partial_i B),\\
P_0&=-pe^{A}(1- e^{B-A} n\partial {B}).
\end{align}

\subsection{Time evolution of the photon energy momentum tensor}

In order to elucidate a concrete collision process, we start with constructing the photon energy momentum tensor from the phase space distribution function $f_{\gamma}$: 
\begin{align}
T^{(\gamma)\mu\nu} &\equiv 2\int \frac{d^4P}{\sqrt{-g}(2\pi)^4}2\pi\delta(P_\alpha P^\alpha)\theta(P^0)2P^\mu P^{\nu} f_\gamma, \label{def:EMT}
\end{align}
where $\theta$ is a step function, $P$'s in this expression are
conjugate momenta $P_\mu$, and $\alpha$ implies a fluid component.
Then the covariant derivative of Eq.~(\ref{def:EMT}) is given by
\begin{align}
\nabla_\mu T^{(\gamma)\mu}{}_{\nu}=2 \int \frac{d^3 P}{\sqrt{-g}(2\pi)^3P^0} P_{\nu} \frac{df_\gamma}{d\lambda},\label{div:emt:f}
\end{align}
where $\lambda$ is an affine parameter and $P^0=d\eta/d\lambda$.
Under the non canonical coordinate transformation $P_i \to p^{\bar a}$
\begin{align}
P_i=g_{ij} e^{j}{}_{\bar a}\frac{p^{\bar a}}{\bar a},
\end{align}
the Jacobian is transformed as
\begin{align}
|g_{ij} e^{j}{}_{\bar a}a^{-1}|=e^{3D}.
\end{align}
Then, the three dimensional volume element in momentum space can be expressed as 
\begin{align}
d^3P\equiv
dP_1dP_2dP_3 = e^{3D}p^2 dp d\mathbf n,
\end{align} 
in terms of the momentum in the local conformal Minkowski frame.
Using the above expression, Eq.~(\ref{div:emt:f}) yields

\begin{align}
\nabla_\mu T^{(\gamma)\mu}{}_{0}=-\frac{2}{a^{4}} \int \frac{p^{2} dp d\mathbf n}{(2\pi)^3} p(1 - n\partial {B}+\cdots ) \frac{df_\gamma}{d\eta},\label{div:emt:f:loc}
\end{align}
where dots imply second order corrections.
The integrand of Eq.~(\ref{div:emt:f:loc}) is directly related to the collision process through the Boltzmann equation:

\begin{align}
\frac{df_\gamma}{d\eta}=\mathcal C[f_{\gamma},\cdots],
\end{align}
where the dots imply the distribution functions of the fluids which
interact with the photons.
When we consider the weak Compton scattering up to second order, a solution to the above Boltzmann equation can written as the superposition of a local blackbody and the spectral $y$ distortion.
In this case, the collision term can be decomposed into the following form~\cite{Ota:2016esq} 
\begin{align}
\mathcal C[f]=\mathcal A \mathcal G(p) +\mathcal B\mathcal Y(p),\label{col:exp}
\end{align}
where we have also introduced
\begin{align}
\mathcal G(p)&\equiv \left(-p\frac{\partial }{\partial p}\right)f^{(0)}(p),\\
\mathcal Y(p)&\equiv \left(-p\frac{\partial }{\partial p}\right)^{2}f^{(0)}(p)- 3\mathcal G(p),
\end{align}
with $f^{(0)}(p) \equiv (e^{p/T_{\rm rf}}-1)^{-1}.$
$p$ is the local frame comoving momentum defined in Eq.~(\ref{pdef}), and $T_{\rm rf}$ is a (constant) comoving temperature of reference blackbody whose number density and energy density are defined as
\begin{align}
N_{\gamma \rm rf}&=2\int\frac{p^{2}dp}{2\pi^{2}}f^{(0)},\\
\rho_{\gamma \rm rf}&=2\int\frac{p^{2}dp}{2\pi^{2}}pf^{(0)}.
\end{align}

We can show that the isotropic component of $\mathcal A$ is zero from the fact that the weak Compton scattering does not change the number of photons. 
Here we introduce the following number density flux
\begin{align}
N^\mu_\gamma\equiv 2\int \frac{d^4P}{\sqrt{-g}(2\pi)^4}2\pi\delta(P_\alpha P^\alpha)\theta(P^0)2P^\mu f_\gamma. \label{def:Nflux}
\end{align}
The covariant derivative of the number flux can be calculated as
\begin{align}
\nabla_\mu N_\gamma^\mu=2\int \frac{d^3P}{\sqrt{-g}(2\pi)^3P^0}\frac{d f_\gamma}{d\lambda}.\label{tochu:covN}
\end{align}
Then, substituting Eqs.~(\ref{col:exp}) into (\ref{tochu:covN}), we
obtain
\begin{align}
\nabla_\mu N_\gamma^\mu=3N_{\gamma \rm rf}\frac{1}{e^{A}a^4}\int \frac{d\mathbf n}{4\pi}\mathcal A=0,
\end{align}
where we have used 
\begin{align}
2\int\frac{p^{2}dp }{2\pi^{2}}\mathcal G &= 3N_{\gamma \rm rf}, \\
2\int\frac{p^{2}dp }{2\pi^{2}}\mathcal Y &= 0.
\end{align}
On the other hand, the dipole component of $\mathcal A$ is not zero.
In our notation, the dipole component of $\mathcal A$ and the monopole component of $\mathcal B$ are written as~\cite{Pitrou:2009bc,Naruko:2013aaa,Chluba:2012gq,Ota:2016esq}
\begin{align}
\int \frac{d\mathbf n}{4\pi}\mathbf n\mathcal A &=\frac 13n_{\rm e} \sigma_{\rm T}a\hat \partial (v+3i\Theta_{1})+\cdots \label{dip:A} \\
\int \frac{d \mathbf n}{4\pi}\mathcal B&=\frac13n_{\rm e} \sigma_{\rm T}a\hat \partial v\hat\partial (v+3i\Theta_{1})\label{Monopo:B},
\end{align}
where the dots represent the second order corrections, and $\hat \partial$
corresponds to $i\mathbf k/|\mathbf k|$ in Fourier space~\footnote{In
Ref.~\cite{Ota:2016esq}, the angular dependence was not properly
treated, and $\hat \partial$ was dropped.}.  $v=|\mathbf v|$ is the
magnitude of the velocity of the baryon fluid, and $\Theta_{1}$ is the
dipole component of the photon temperature perturbations.  $n_{\rm e}$
is the electron density, $\sigma_{\rm T}$ is the Thomson scattering
cross section, and $a$ is a scale factor.  Using
Eqs.~(\ref{div:emt:f:loc}), (\ref{col:exp}), (\ref{dip:A}) and
(\ref{Monopo:B}), we find
\begin{align}
\nabla_\mu T^{(\gamma )\mu}{}_{0}=-
\frac{4}{3a^{4}}\rho_{\gamma,{\rm rf}}n_{\rm e} \sigma_{\rm T}a(\hat\partial v-\partial B)\hat\partial (v+3i\Theta_{1}),\label{res:cons:emt}
\end{align}
where we have used
\begin{align}
2\int\frac{p^{2}dp p}{2\pi^{2}}\mathcal G&=4\rho_{\gamma \rm rf},\\
2\int\frac{p^{2}dp p}{2\pi^{2}}\mathcal Y&=4\rho_{\gamma \rm rf}.
\end{align}

We are now ready to discuss the superhorizon evolution of the
isocurvature perturbations in the presence of heat conduction between
electrons and photons. 
From Eqs.~(\ref{emtdiv0com3}), (\ref{cons:grad:second}), and~(\ref{res:cons:emt}), we find
\begin{align}
\xi_\gamma'=& \frac{1}{3} n_{\rm e} \sigma_{\rm T}a(\hat\partial v-\partial B)\hat\partial (v+3i\Theta_{1}),
\label{emtdiv0com4}\\
\xi_b'=&
-\frac{1}{3R} n_{\rm e} \sigma_{\rm T}a(\hat\partial v-\partial B)\hat\partial (v+3i\Theta_{1}),
\label{emtdiv0com5}\\
\xi_c'=&0,
\label{emtdiv0com6}
\end{align}
where $R=3\rho_{b}/4\rho_{\gamma}=3a\rho_{b,{\rm rf}}/4\rho_{\gamma,{\rm rf}}$, and we used Eq.~(\ref{emtdiv0com3}) for the baryon fluid with
\begin{align}
\nabla_\mu T^{(\gamma )\mu}{}_{0}+ \nabla_\mu T^{(b)\mu}{}_{0}=0.
\end{align}
Then time derivatives of the isocurvature perturbations defined with Eq.~(\ref{def:iso:xi}) become
\begin{align}
S_{b\gamma }'&=-\frac{(1+R)}{R} n_{\rm e} \sigma_{\rm T}a(\hat\partial v-\partial B)\hat\partial (v+3i\Theta_{1}),\label{Sgabbabprime}\\
S_{c\gamma }'&=- n_{\rm e} \sigma_{\rm T}a(\hat\partial v-\partial B)\hat\partial (v+3i\Theta_{1}).\label{Sgabbabprimec}
\end{align}
These expressions imply that the heat conduction from electron fluid is
responsible for the change of the total photon energy while the friction
heat from the intrinsic photon shear viscosity $\Theta_2$ is not. This
is because the friction heat from the photon anisotropic stress does not
increase the net energy in a photon system as long as we deal with
background and perturbations as a whole system.  Some confusion may
occur if one separates the background and perturbations as done in the
previous literatures, in which energy transfers from perturbations to
the background are discussed.  In response to Eq.~(\ref{res:cons:emt}),
the energy momentum conservation for baryons should be also broken while
those of the total fluids and the other dark sectors remain conserved.
Note that these expressions are independent of the gauge
choice~(\ref{mygauge}) since Eq.~(\ref{gt:q}) for the baryons and the
photons are written as
\begin{align}
v&\to \tilde v=v+k\alpha,\\
\Theta_{1}&\to \tilde \Theta_{1}=\Theta_{1}+\frac{ik}{3}\alpha.
\end{align}
\\

\subsection{Role of the primordial non Gaussianity}

Eqs.~(\ref{Sgabbabprime}) and (\ref{Sgabbabprimec}) imply that the observed isocurvature perturbations are superposition of the primordial isocurvature and the secondary isocurvature.
Suppose we only have the adiabatic perturbations at the beginning, the Fourier space isocurvature perturbations are simply given as
\begin{align}
S_{\alpha \gamma,\mathbf k} &= \int \frac{d^{3}k_{1}d^{3}{k_{2}}}{(2\pi)^{6}}(2\pi)^{3}\delta^{(3)}(\mathbf k_{1}+\mathbf k_{2}-\mathbf k) \mathcal S_{\alpha}(\mathbf k_{1},\mathbf k_{2})\zeta_{\mathbf k_{1}}\zeta_{\mathbf k_{2}},\label{So:def}
\end{align}
Here, the transfer functions in Fourier space are introduced as
\begin{align}
\mathcal S_{\alpha }(\mathbf k_{1},\mathbf k_{2})= \hat k_{1}\cdot \hat k_{2} \int d\eta  w_{\alpha} n_{\rm e}\sigma_{\rm T}a[v(k_{1})-k_{1} B(k_{1})][v(k_{2})+3i\Theta_{1}(k_{2})],
\end{align}
where $w_{b}=(1+R)/R$, $w_{c}=1$.
On the other hand, the statistics of the adiabatic perturbations in the Fourier spaces are written as
\begin{align}
\langle \zeta_{\mathbf k_1}\zeta_{\mathbf k_2}\rangle
&=(2\pi)^3\delta^{(3)}\left[\sum_{i=1}^2 \mathbf k_i\right]P_\zeta(k_1),\label{power:zeta}\\
\langle \zeta_{\mathbf k_1}\zeta_{\mathbf k_2}\zeta_{\mathbf k_3}\rangle 
&=(2\pi)^3\delta^{(3)}\left[\sum_{i=1}^3 \mathbf k_i\right]B_\zeta(\mathbf k_1,\mathbf k_2,\mathbf k_3),\label{bis:zeta}\\
\langle \zeta_{\mathbf k_1}\zeta_{\mathbf k_2}\zeta_{\mathbf k_3}\zeta_{\mathbf k_4}\rangle&=(2\pi)^3\delta^{(3)}\left[\sum_{i=1}^4 \mathbf k_i\right]T_\zeta(\mathbf k_1,\mathbf k_2,\mathbf k_3,\mathbf k_4).\label{tri:zeta}
\end{align}
Then, the cross correlations with the adiabatic perturbations and the auto correlations become

\begin{align}
\langle S_{\alpha\gamma ,\mathbf k}\zeta_{\mathbf k'} \rangle &=(2\pi)^{3}\delta(\mathbf k+\mathbf k') P_{\alpha \zeta}(\mathbf k),\\
\langle S_{\alpha \gamma ,\mathbf k}S_{\beta \gamma\mathbf k'} \rangle &=(2\pi)^{3}\delta(\mathbf k+\mathbf k') P_{\alpha \beta}(\mathbf k),
\end{align}
where the powerspectra are calculated as
\begin{align}
P_{\alpha \zeta}&=\int \frac{d^{3}k_{1}}{(2\pi)^{3}} \mathcal S_{\alpha}(\mathbf k_{1},\mathbf k- \mathbf k_{1}) B_{\zeta}(\mathbf k_{1},\mathbf k-\mathbf k_{1},\mathbf k),\\
P_{\alpha \beta}&=\prod_{i=\alpha,\beta}\left[\int \frac{d^{3}k^{(i)}_{1}}{(2\pi)^{3}}\mathcal S_{i}(\mathbf k^{(i)}_{1},\mathbf k- \mathbf k^{(i)}_{1}) \right]  T_{\zeta}(\mathbf k^{(\alpha)}_{1},\mathbf k-\mathbf k^{(\alpha)}_{1},\mathbf k^{(\beta)}_{1},\mathbf k-\mathbf k^{(\beta)}_{1}).
\end{align}
The scale dependences of the secondary powerspectra depend on the shape of the primordial non Gaussianity.
As an example, consider the local forms of bispectra and trispectra:
\begin{align}
&B_\zeta(\mathbf k_1,\mathbf k_2,\mathbf k_3)=\frac65f^{\rm loc.}_{\rm NL}\left[P_{\zeta}(k_1)P_{\zeta}(k_2) + (\text{2 perms.}) \right],\label{def:fnl}\\
&T_\zeta(\mathbf k_1,\mathbf k_2,\mathbf k_3,\mathbf k_4)=\tau^{\rm loc.}_{\rm NL}\left[P_{\zeta}(k_1)P_{\zeta}(k_2)P_{\zeta}(|\mathbf k_1+\mathbf k_3|) + (\text{11 perms.}) \right]\label{def:tnl},
\end{align}
where we have omitted terms proportional to $g^{\rm loc.}_{\rm NL}$ for simplicity.
Then the dominant contributions become

\begin{align}
P_{\alpha \zeta}&\approx \frac{12}{5}f^{\rm loc.}_{\rm NL}P_{\zeta}(k)\times \int \frac{d^{3}k_{1}}{(2\pi)^{3}} \mathcal S_{\alpha}(\mathbf k_{1},-\mathbf k_{1})P_{\zeta}(k_{1}),\\
P_{\alpha \beta}&\approx 4\tau^{\rm loc.}_{\rm NL}P_{\zeta}(k) \times \prod_{i=\alpha,\beta}\left[\int \frac{d^{3}k^{(i)}_{1}}{(2\pi)^{3}}S_{i}(\mathbf k^{(i)}_{1},-\mathbf k^{(i)}_{1})P_{\zeta}(k^{(i)}_{1}) \right].
\end{align}
Thus, the powerspectra of the secondary isocurvature perturbations are the same form with the linear isocurvature powerspectrum.
The disconnected part of the trispectrum leads to the following contribution for $\mathbf k\neq 0$ and $\mathbf k'\neq 0$:

\begin{align}
P^{(d)}_{\alpha \beta}&\approx \int \frac{d^{3}k_{1}}{(2\pi)^{3}} \prod_{i=\alpha,\beta}\ \mathcal S_{i}(\eta^{i},\mathbf k_{1},-\mathbf k_{1}) P_{\zeta}(k_{1})P_{\zeta}(k_{1}).
\end{align}

Then we obtain $P^{(d)}_{\alpha \beta}\approx{\rm const.}$ for the disconnected trispectrum.
This suggests the spectral index is 4, and the powerspectrum is mainly enhanced on scales where the physical process occurs.
In other words, the Gaussian fluctuations cannot produce the superhorizon isocurvature modes.

\section{Generation of Entropy perturbations}
\label{sec:entropy}

Besides the conserved quantity~(\ref{cons:grad:second}), one may wonder if we could introduce the similar quantities by using the entropy flux.
In this section, we introduce the secondary entropy perturbations, which are not identified with the isocurvature perturbations if we consider non equilibrium universe during recombination.

\subsection{Entropy flux non conservation}

Suppose the universe is out of equilibrium states, the standard thermodynamic relation among the entropy density, the
energy density and pressure is not applicable. Instead, we
introduce the (Shannon) entropy flux, which is defined in terms of a logarithm of the number of
states~\cite{Khatri:2012rt},
\begin{align}
S^\mu_\gamma&\equiv 2\int
 \frac{d^4P}{\sqrt{-g}(2\pi)^4}2\pi\delta(P_\alpha
 P^\alpha)\theta(P^0)2 P^\mu  \mathcal F \label{def:shannon},\\
\mathcal F&\equiv \left[(f_\gamma+1)\ln(f_\gamma+1)-f_\gamma\ln f_\gamma \right].\label{def:calF}
\end{align}
Note that this definition
reproduces the thermodynamic entropy density for the Planck
distribution. The covariant divergence of this entropy flux can be
calculated as
\begin{align}
\nabla_\mu S^\mu_\gamma =2\int \frac{d^3P}{\sqrt{-g}(2\pi)^3P^0}\frac{d
 \mathcal F}{d\lambda}.\label{tochu:divS}
\end{align}

A solution to the Boltzmann equation with the weak Compton collision process can be written as a superposition of the local blackbody and the spectral $y$ distortion up to the second order in the primordial fluctuations~\cite{Pitrou:2009bc,Naruko:2013aaa}.
Such an ansatz can be expanded as follows:
\begin{align}
f_{\gamma}=&f^{(0)}(p)+\left[\Theta +\frac32\Theta^2\right] \mathcal G(p) + \left[\frac12\Theta^2+y\right]\mathcal Y(p),\label{def:non-thermal_anz}
\end{align}
where $\Theta=\Theta^{(1)}+\Theta^{(2)}$ and $y=y^{(2)}$ are the temperature perturbation and spectral $y$ distortion, respectively.
Then, Eq.~(\ref{tochu:divS}) vanishes at zeroth and
first orders of the perturbations, but there exist non-zero
contributions at second order, which is manifest from the following
expression,
\begin{align}
\frac{1}{P^0}\frac{d \mathcal F}{d\lambda} &=
\frac{d \mathcal F}{d\eta}=\frac{p}{T_{\rm rf}}\left[ \left(1-\Theta\right)\mathcal A\mathcal G+\mathcal B\mathcal Y\right].\label{calc:F2}
\end{align}
Here we have replaced the Liouville term with the collision terms by
using the Boltzmann equation.
Using the Boltzmann equation for the $y$ distortion~\cite{Pitrou:2009bc,Naruko:2013aaa,Chluba:2012gq,Ota:2016esq},
\begin{align}
y'&=\mathcal B-\Theta\mathcal A,
\label{y:eq}
\end{align}
with Eqs.~(\ref{tochu:divS}), and (\ref{calc:F2}), we find
\begin{align}
\nabla_\mu S^\mu_\gamma= \frac{4\pi^2}{15 a}\left(\frac{T_{\rm rf}}{a}\right)^3 y'_0.\label{eq:14}
\end{align}
Thus, entropy increases with the generation of the spectral $y$
distortion. The physical entropy density can be defined as
$S_\gamma\equiv - n_{\mu} S^{\mu}_\gamma$ with $ n_{\mu}\equiv
{\nabla_\mu \eta}(-\nabla_\nu\eta \nabla^\nu\eta)^{-\frac12}$ being the
normalized 1-form orthogonal to a constant $\eta$ hypersurface. One may
wonder if Eq.~(\ref{eq:14}) can also be derived from the standard
thermodynamic relation,
\begin{align}
\frac{dS_\gamma}{dt}=\frac{1}{T}\frac{dQ}{dt},\label{defthent}
\end{align}
where $Q$ is thermodynamical heat. 
What we found is not a reinterpretation of this relation because we identify ``heat'' for the photon baryon fluid in the presence of non-equilibrium effect; thermodynamic arguments are not applicable
to.
Thus, the generation of $y$ distortion is not directly
identified with the entropy perturbation production without a kinetic description based
on the Boltzmann equation. \\

\subsection{Entropy perturbations at second order}

We are now ready to introduce a quantity
\begin{align}
\zeta^{(S)}_\gamma\equiv D+\frac{A}{3}+\frac13\ln \left(
 \frac{S^0}{S^0_{\rm rf}}\right),\label{zeta_S}
\end{align}
where $S^{0}_{\rm rf}= 4\pi^2T_{\rm rf}^3/(45a^4)$.
This quantity is conserved as long as entropy flux conserves at leading order of the gradient expansion.
Let us check this statement by considering the covariant derivative of the entropy flux:
\begin{align}
\nabla_\mu S^\mu_\gamma=\partial_\mu S^{\mu}+S^\mu\partial_\mu\ln \sqrt{-g}.
\end{align}
Dropping a gradient term $\partial_i S^i$, we find
\begin{align}
\zeta_\gamma ^{(S)}{}'= -\frac{S^i}{3S^0}\partial_i(A+3D)+y'_0,
\end{align}
where we have used Eq.~(\ref{eq:14}). 
The first term represents a
volume effect, which is manifest only when we take into account the next
leading order of the gradient expansion.  $\zeta^{(S)}_{\gamma}$ is
conserved even at second order when the scattering is negligible, but
only if we move on to $A+3D=0$ gauge, where the volume element does not
fluctuate.  However, note that gauge is not completely fixed on this
slice. The second term arises as a result of the entropy production,
which, in this paper, we should keep since the imperfectness of a fluid
on subhorizon scales could be non negligible due to convolutions.\\

The entropy density is not necessarily proportional to the number density if both of
them are evaluated for a non-equilibrium state. In our case, its
discrepancy is expressed in terms of $y$ distortion, which characterizes
the deviation from the thermodynamic system.
The curvature perturbations on the uniform number density slice can be also defined through the same procedures with the entropy:
\begin{align}
\zeta^{(N)}_\gamma\equiv D+\frac{A}{3}+\frac13\ln \left(
 \frac{N^0}{N^0_{\rm rf}}\right),
\end{align}
where $N^{0}_{\rm rf}= 2\zeta(3) T_{\rm rf}^3/(\pi^2 a^4)$.
Using the number flux conservation laws and dropping $\partial_i N^i$, we find
\begin{align}
\zeta_\gamma ^{(N)}{}'=-\frac{N^i}{3N^0}\partial_i(A+3D).
\end{align}
Thus $\zeta_\gamma ^{(N)}$ is also a conserved quantity if we have the number conservation law and take the leading order of the gradient expansion.
Note that $\zeta^{(N)}_\gamma$ is also conserved in $A+3D=0$ gauge even at second order without truncating the higher order gradient corrections.
\\

Now let us consider the following isocurvature perturbations:
\begin{align}
S^{(NS)}_{\alpha \gamma} \equiv \zeta^{(N)}_{\alpha} - \zeta^{(S)}_{\gamma}.\label{def:iso:NS}
\end{align}
This is a covariant extension of $\delta \left(N_{\alpha}/S_{\gamma}\right)$ at nonlinear order.
It should be noticed that the following relation 
\begin{align}
\frac{N^i}{N^0}=\frac{S^i}{S^0},
\end{align}
applies at linear order even for the present case since the spectral distortion is a second order
effect.
Then we obtain 
\begin{align}
S^{(NS)'}_{\alpha \gamma} = -y_{0}'.
\end{align}
Thus, the entropy perturbations are also conserved quantity in the presence of gradient terms if the photon entropy flux and $\alpha$-fluid number density flux are conserved.
We may also consider isocurvature perturbations defined as
\begin{align}
S^{(N)}_{\alpha \gamma}\equiv 3(\zeta^{(N)}_\alpha-\zeta^{(N)}_\gamma),\label{def:photoniso:nn}
\end{align}
which are conserved if each number density flux are conserved.

The above
discordance between Eqs.~(\ref{def:iso:NS}) and (\ref{def:photoniso:nn}) motivates us to newly
define the \textit{photon isocurvature perturbation} as a fluctuation of a fraction between the photon number density and the photon entropy density
\begin{align}
S^{(NS)}_{\gamma\gamma}=-3y_0.
\end{align}
This is nothing but the spectral $y$ distortion.
For the chemical equilibrium period in the early universe where $y$ distortion is erased, it is obvious that $S^{(N-S)}_{\alpha\gamma}=S^{(N)}_{\alpha\gamma}=S_{\alpha\gamma}$ due to thermodynamic relations.
\\

Thus, Eqs.~(\ref{def:iso:NS}) and (\ref{def:photoniso:nn}) can be also
defined as superhorizon conserved quantities without scattering
processes.  However, in contrast to the conservation laws of energy
momentum tensor, the conservation laws for the number flux and the
entropy flux are not necessarily established in the whole cosmic
history.  Therefore, Eq.~(\ref{def:iso:xi}) is much more important than
the others.

\section{Conclusions}\label{conclusion}

In this paper, we revisited the two assumptions for the conservation
laws of the superhorizon isocurvature perturbations: the negligibility
of the gradient terms and the energy conservation laws for the component
fluids.  We pointed out that the second order gradient terms are not
necessarily dropped even if we consider the long wavelength modes.
Then, we have introduced new second order quantities, which are
conserved even in the presence of gradient terms if there are no
non-adiabatic pressure perturbations. It should be noticed that they
coincide with the curvature perturbations on the uniform density slice
only when we can ignore the gradient terms. The total energy momentum
tensor is always conserved, but that for each component fluid is not
necessarily conserved.  As such an example, we discuss the weak Compton
scattering that transfers the energy between the photons and baryons.
We found that the secondary isocurvature perturbations are generated due
to this energy transfer.  The powerspectra of secondary isocurvature
perturbations become scale invariant if we consider the local form of
the primordial tri- and bispectrum.  On the other hand, the disconnected
part of the trispectrum only produces the isocurvature perturbations on
scales where the actual physical process occurs.  We also commented on
the entropy perturbations, which are usually equivalent to the
isocurvature perturbations in thermal equilibrium states. However, in
our case, we cannot identify these two quantities when the universe is
dominated by the weak Compton scattering and is not in thermal
equilibrium. We found that the entropy perturbations can be understood
in terms of the spectral $y$ distortion, which is a non thermal
deviation from the blackbody spectrum produced in the weak Compton
scattering dominated universe.

The new quantity $\xi$ we have introduced in this paper is still gauge
dependent. However, it should be noticed that we can always define the
gauge invariant quantities recursively even at nonlinear order as
pointed out in Ref.~\cite{Nakamura:2014kza}.  Using this formalism, the
gauge invariant expressions for $\xi$ would be investigated in future
works.  Though we only consider the weak Compton scattering, it would
also be interesting if we consider the similar heat conduction from the
other species such as neutrinos in the earlier epoch.  This would lead
to a new constraint on curvature perturbations with extremely short
wavelength though it requires explicit evaluation for each scattering
process, which is left for our future works.  So far, we have discussed
the late epoch when the universe is in neither kinetic nor chemical
equilibrium.  In the early epoch, the full considerations of the Compton
collision terms are necessary. When there exist relativistic electrons
that can sufficiently transfer the photon energy, local kinetic
equilibrium is expected.  In this case, the $y$ distortion may be
transformed into the $\mu$ distortion, which is defined as chemical
potential of a Bose distribution function.  In the earlier epoch, the
number changing process such as the double Compton effects,
Bremsstrahlung or pair annihilation are also non-negligible.  They
adjust the number density and erase the spectral distortions so as to
realize chemical equilibrium.  Referring to Eqs.~(\ref{tochu:covN}) and
(\ref{div:emt:f}), such violation of photon number density conservation
would break photon energy conservation as well. Then, secondary
isocurvature perturbations might be additionally generated on
superhorizon scales, but further study is necessary to make a clearer
statement.


\begin{acknowledgments}
We would like to thank Misao Sasaki and Atsushi Naruko for useful
discussion on conservation of isocurvature perturbations on superhorizon
scales.  The authors are grateful to Kouji Nakamura and Karim Malik for
helpful discussions.  We also would like to thank Rampei Kimura for
careful reading of our manuscript.  This work was supported in part by
JSPS Grant-in-Aid for PD Fellows (A.O.), JSPS Grant-in-Aid for
Scientific Research Nos.~25287054 (M.Y.)  and 26610062 (M.Y.), MEXT
KAKENHI for Scientific Research on Innovative Areas ``Cosmic
Acceleration'' No. 15H05888 (M.Y.).

\end{acknowledgments}

\bibliography{bib}{}
\bibliographystyle{unsrt}

\end{document}